\documentclass[reprint, amsmath,amssymb, aps, prl, longbibliography, lengthcheck]{revtex4-1}

\usepackage{graphicx}
\usepackage{dcolumn}
\usepackage{bm}
\usepackage{hyperref}
\usepackage{epsfig}

\begin{document}

\title{Sympathetic cooling of rovibrationally state-selected molecular ions}

\author{Xin Tong}
\author{Alexander H. Winney}
\author{Stefan Willitsch}
\email{stefan.willitsch@unibas.ch}
\affiliation{Department of Chemistry, University of Basel, Klingelbergstrasse 80, 4056 Basel, Switzerland}

\date{\today}

\begin{abstract}
We present a new method for the generation of rotationally and vibrationally state-selected, translationally cold molecular ions in ion traps. Our technique is based on the state-selective threshold photoionization of neutral molecules followed by sympathetic cooling of the resulting ions with laser-cooled calcium ions. Using N$_2^+$ ions as a test system, we achieve $>90$~\% selectivity in the preparation of the ground rovibrational level and state lifetimes on the order of 15 minutes limited by collisions with background-gas molecules. The technique can be employed to produce a wide range of apolar and polar molecular ions in the ground and excited rovibrational states. Our approach opens up new perspectives for cold  quantum-controlled ion-molecule-collision studies, frequency-metrology experiments with state-selected molecular ions and molecular-ion qubits.
\end{abstract}

\maketitle

Recent advances in the preparation of translationally cold neutral molecules and molecular ions \cite{carr09a} have opened up perspectives for studies of molecular collisions and chemical reactions at extremely low temperatures \cite{willitsch08a, knoop10a, ospelkaus10b},  ultrahigh-precision molecular spectroscopy \cite{schiller05a} and novel schemes for quantum-information processing \cite{demille02a}. Such experiments not only require precise control over the translational motion of the molecules, but also over their internal, in particular rotational-vibrational, quantum state. The preparation of fully quantum-state-selected ultracold molecules and molecular ions constituted one of the major challenges in the field over the past decade and has only very recently been achieved for neutral diatomics synthesized from ultracold atoms (see, e.g., Refs. \cite{carr09a, ospelkaus10b, knoop10a} and references therein).

Translationally cold molecular ions, on the other hand, are conventionally produced from "hot" samples by sympathetic cooling using the Coulomb interaction with laser-cooled atomic ions \cite{willitsch08b}. Because low-energy collisions between ions are dominated by the Coulomb interaction which does not couple to the internal degrees of freedom, sympathetically-cooled ions exhibit broad distributions of rotational-state populations \cite{bertelsen06a,koelemeij07a}. In such translationally cold, but internally warm samples population can be accumulated in the rotational ground state using optical pumping schemes as demonstrated in two recent studies by Staanum et al. \cite{staanum10a} and Schneider et al. \cite{schneider10a}. In their experiments, continuous excitation of selected rovibrational transitions in combination with population redistribution aided by black-body radiation (BBR) enabled to increase the ground-state population to 37\% in MgH$^+$ \cite{staanum10a} and 78\% in HD$^+$ \cite{schneider10a}. Although these values are about an order of magnitude higher than the corresponding thermal populations at room temperature, the state preparation is not complete which reflects the challenges associated with optical pumping in systems with a large number of simultaneously populated levels. Moreover, because the schemes used thus far rely on population transfer via dipole-allowed transitions, they cannot be applied to fundamental apolar ions such as H$_2^+$, N$_2^+$ and O$_2^+$.

In the present letter, we demonstrate a complementary method for the production of rovibrationally state-selected, translationally cold molecular ions which circumvents the problems associated with optical pumping applied to a broad distribution of state populations. Our approach is based on the initial generation of the ions in the desired rovibrational quantum state using state-selective threshold-photoionization \cite{willitsch05b, mackenzie95a} immediately followed by sympathetic cooling of their translational motion. Using N$_2^+$ as a test system, we achieve a population of $93\pm11$~\% in the rotational ground state after sympathetic cooling. 

\begin{figure}[t]
\epsfig{file=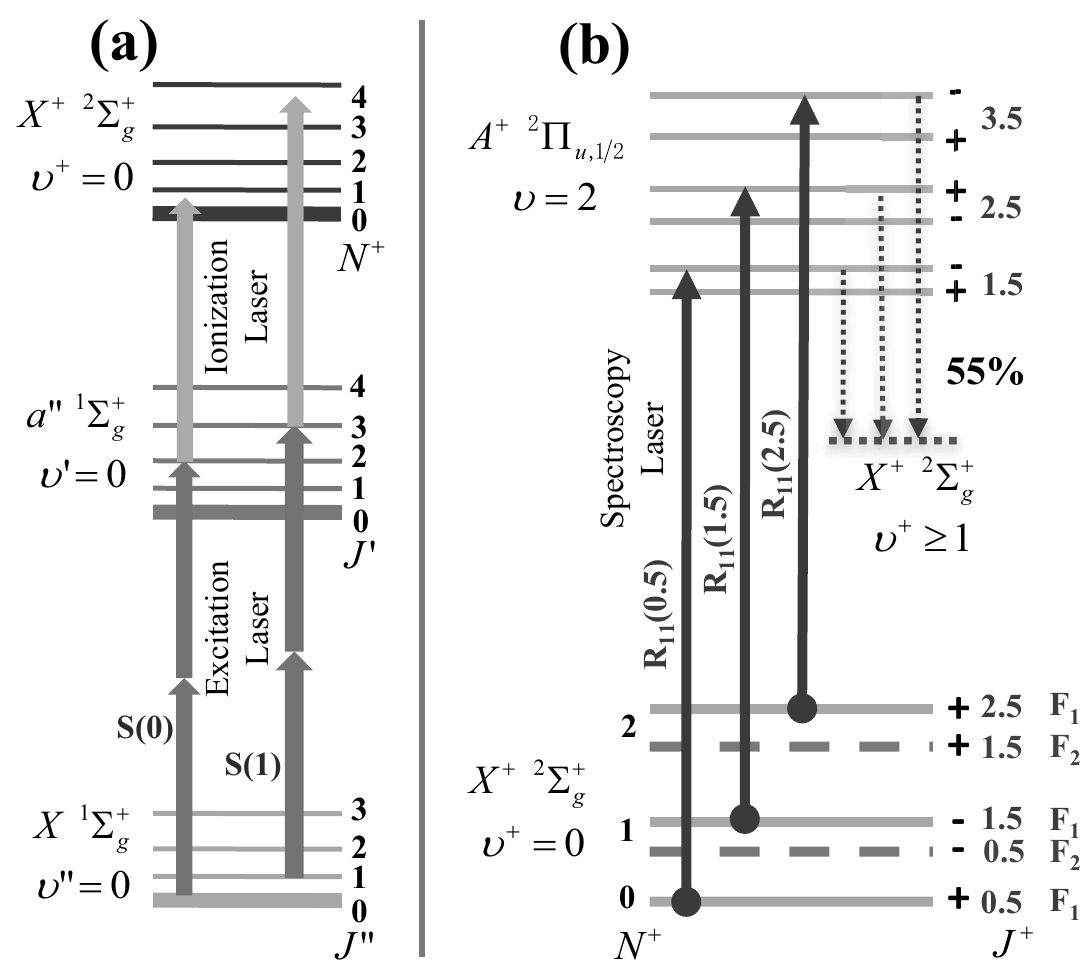,width=0.9\columnwidth}
\caption{\label{exc} (a) State-selective $[2+1']$ resonance-enhanced threshold-photoionization sequence for the generation of N$_2^+$ ions in the $N^+=0$ and 3 rotational states. (b) Excitation scheme for laser-induced charge transfer (LICT) used to probe the population in spin-rotational levels of N$_2^+$. See text for details.}
\end{figure}

Rovibrationally state-selected N$_2^+$ ions were produced inside a linear radiofrequency (RF) ion trap \cite{willitsch08b} by resonance-enhanced $[2+1']$-photon threshold photoionization via selected rotational levels of the $a''~^1\Sigma^+_g,~v'=0$ excited state of N$_2$. Here, $v$ stands for the vibrational quantum number and the superscripts $'','$ and $^+$ denote the neutral ground, neutral excited and cationic ground electronic states. Capitalizing on a rotational propensity rule strongly favoring $\Delta N=N^+-J'=0$ transitions (with weaker $\Delta N=\pm2$ satellites) for ionization out of the $a''~^1\Sigma^+_g$ intermediate state \cite{mackenzie95a}, rovibrational state selection in the cation was achieved by photoionizing slightly above the lowest rotational ionization thresholds accessible from a well-defined rotational level in the intermediate state, see Fig. \ref{exc} (a). $N^+$ and $J'$ denote the rotational angular-momentum quantum numbers in the cationic and intermediate states, respectively. The spin-rotation splitting in excited rotational levels of the cationic ground state \cite{wu07a} was not resolved at the bandwidth of the ionization lasers used (0.2 cm$^{-1}$).

Resonance-enhanced $[2+1']$ photoionization was carried out using the frequency-tripled ($\lambda_1\approx202$~nm) and -doubled ($\lambda_2\approx375$~nm) output of two Nd:YAG-pumped pulsed dye lasers. State-selection in the intermediate state was achieved by setting $\lambda_1$ to a well-separated two-photon rovibronic transition from the ground electronic state (see Fig. \ref{exc} (a)) \cite{hanisco91a}. The N$_2^+$ ions were produced inside a linear radiofrequency ion trap \cite{willitsch08b} by focussing both laser beams into a molecular beam of pure nitrogen passing through the trap region (repetition rate 10 Hz, pulse duration 1 ms, diameter $\approx 250~\mu$m). The ion trap was operated at a frequency $\Omega_\text{RF}=2\pi\times3.2$~MHz and a RF amplitude $V_\text{RF,0-p}=120$~V. The operating pressure in the ion trap chamber was maintained at $p=6\times10^{-10}$~mbar. The molecular beam passed the center of the ion trap at a distance of $\approx 400~\mu$m to suppress inelastic collisions between the sympathetically-cooled ions and neutral molecules from the beam. The laser intensities (0.2 mJ/pulse at 202 nm, 1.5 mJ at 375 nm) and focussing conditions were carefully adjusted to suppress the production of spurious ions originating from unselective [2+1] photoionization of nitrogen and background gas molecules. 

\begin{figure}[b]
\epsfig{file=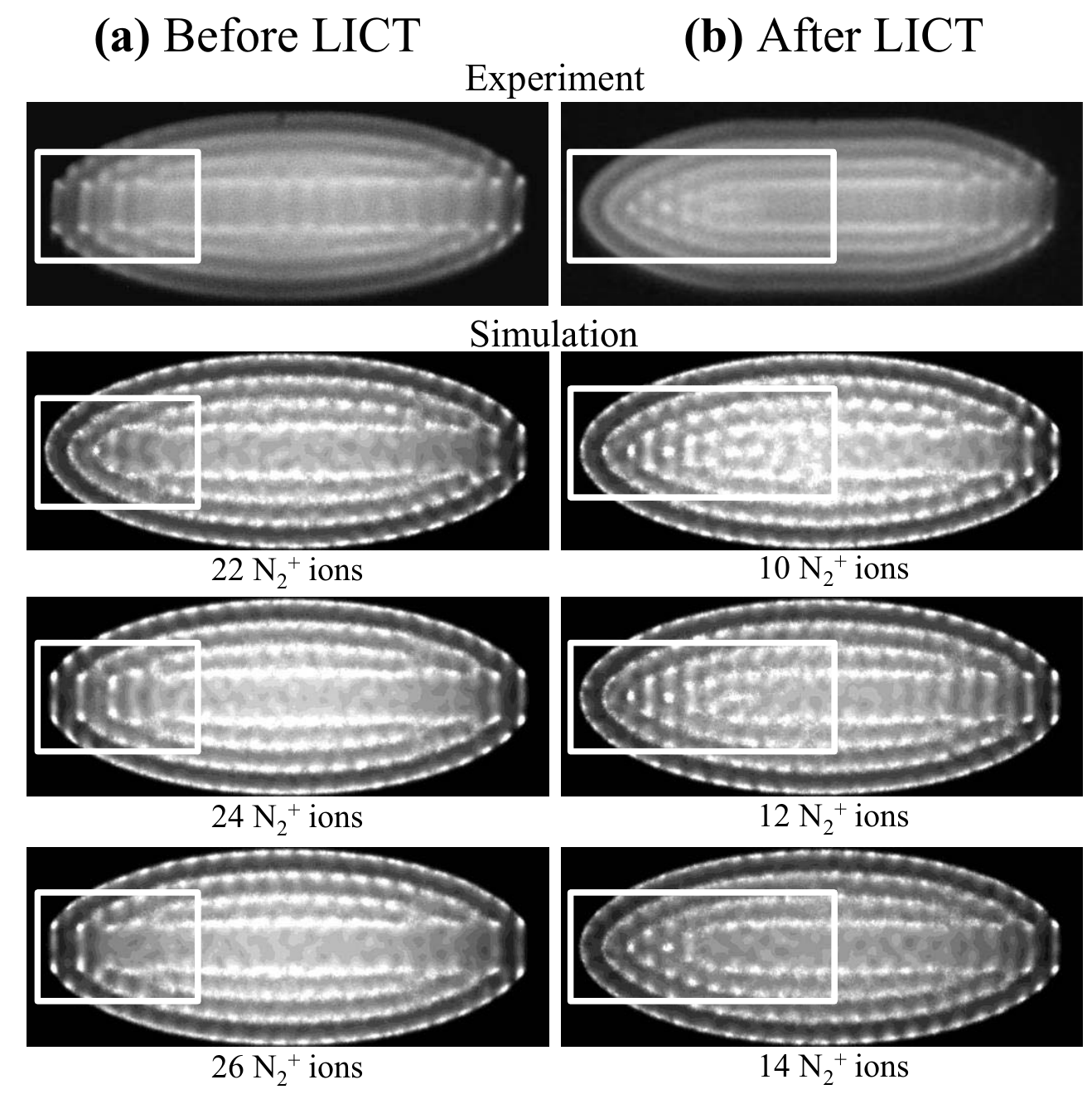,width=0.85\columnwidth}
\caption{\label{num} Determination of the number $N_{\text{N}_2^+}$ of state-selected sympathetically-cooled N$_2^+$ ions by a comparison between experimental and simulated images. The images were obtained (a) before and (b) after laser-induced charge transfer (LICT). The shape of the dark core in the images sensitively depends on $N_{\text{N}_2^+}$, see the highlighted regions to guide the eye.}
\end{figure}

The state-selected N$_2^+$ ions were sympathetically cooled using the Coulomb interaction with simultaneously trapped laser-cooled Ca$^+$ ions (see Ref. \cite{willitsch08b} for details of the laser-cooling setup). The resulting ordered structures of translationally cold ions (conventionally termed "Coulomb crystals" \cite{willitsch08b}) were imaged by collecting the spatially resolved fluorescence of the Ca$^+$ ions using a CCD camera coupled to a microscope. The N$_2^+$ ions do not fluoresce and are only indirectly visible as a dark core in the center of the Coulomb crystals, see Fig. \ref{num} (a). Its asymmetric shape is a consequence of radiation pressure on the Ca$^+$ ions. The precise appearance of the dark core region sensitively depends on the number of sympathetically cooled ions. By carefully comparing the experimental results with theoretical images obtained from molecular dynamics (MD) simulations containing a well-defined number of sympathetically-cooled molecular ions \cite{zhang07a}, the number of N$_2^+$ ions in the experimental images could be determined with an accuracy of $\pm 1$ ion  as demonstrated in Fig. \ref{num}. The comparison with the simulated images also enabled an estimate of the average thermal energies of the ions in the crystals ($E_\text{th}/k_\text{B}=8\pm1$~mK for Ca$^+$ and $16\pm1$~mK for N$_2^+$).

The rotational-state populations of the sympathetically-cooled N$_2^+$ ions were probed using laser-induced charge-transfer (LICT) with Ar atoms \cite{schlemmer99a}. Vibrationally excited N$_2^+$ ions react with Ar according to $\mathrm{N}_2^+ (v^{+}\geq1)\ +\ \mathrm{Ar \rightarrow N_2 \ + \ Ar^+}$. For ions in the vibrational ground state $v^{+}=0$, the reaction is endothermic and thus suppressed under our experimental conditions. In the present experiment, the N$_2^+$ ions were optically excited from selected spin-rotational levels in $v^+=0$ to the $A^+~^2\Pi_{u,1/2},~v=2$ state using 15 mW of diode-laser radiation at a wavelength around 785 nm (see Fig. \ref{exc} (b)) \cite{wu07a, schlemmer99a}. Fluorescent decay of the electronically excited state populates vibrationally excited levels $v^{+}\geq 1$ in the $X^+~^2\Sigma_g^+$ ground electronic state with a probability of 55\% \cite{schlemmer99a} putting an upper limit to the proportion of N$_2^+$ ions which can undergo charge transfer.  Charge transfer was initiated immediately after laser excitation by introducing argon gas at a pressure $p(\text{Ar})=5\times10^{-8}$~mbar into the chamber through a leak valve.

The Ar$^+$ ions produced by LICT remain trapped and are sympathetically cooled. However, the Ar$^+$ ions do not localize in the center of the crystal, but mix freely with the Ca$^+$ ions because of the almost identical masses of the two species. Consequently, the removal of N$_2^+$ ions as a consequence of charge transfer with Ar can directly be observed in the images (see Fig. \ref{num} (b)) and the proportion of removed N$_2^+$ ions represents a measure of the population in the rotational level probed by LICT. 

\begin{figure}[t]
\epsfig{file=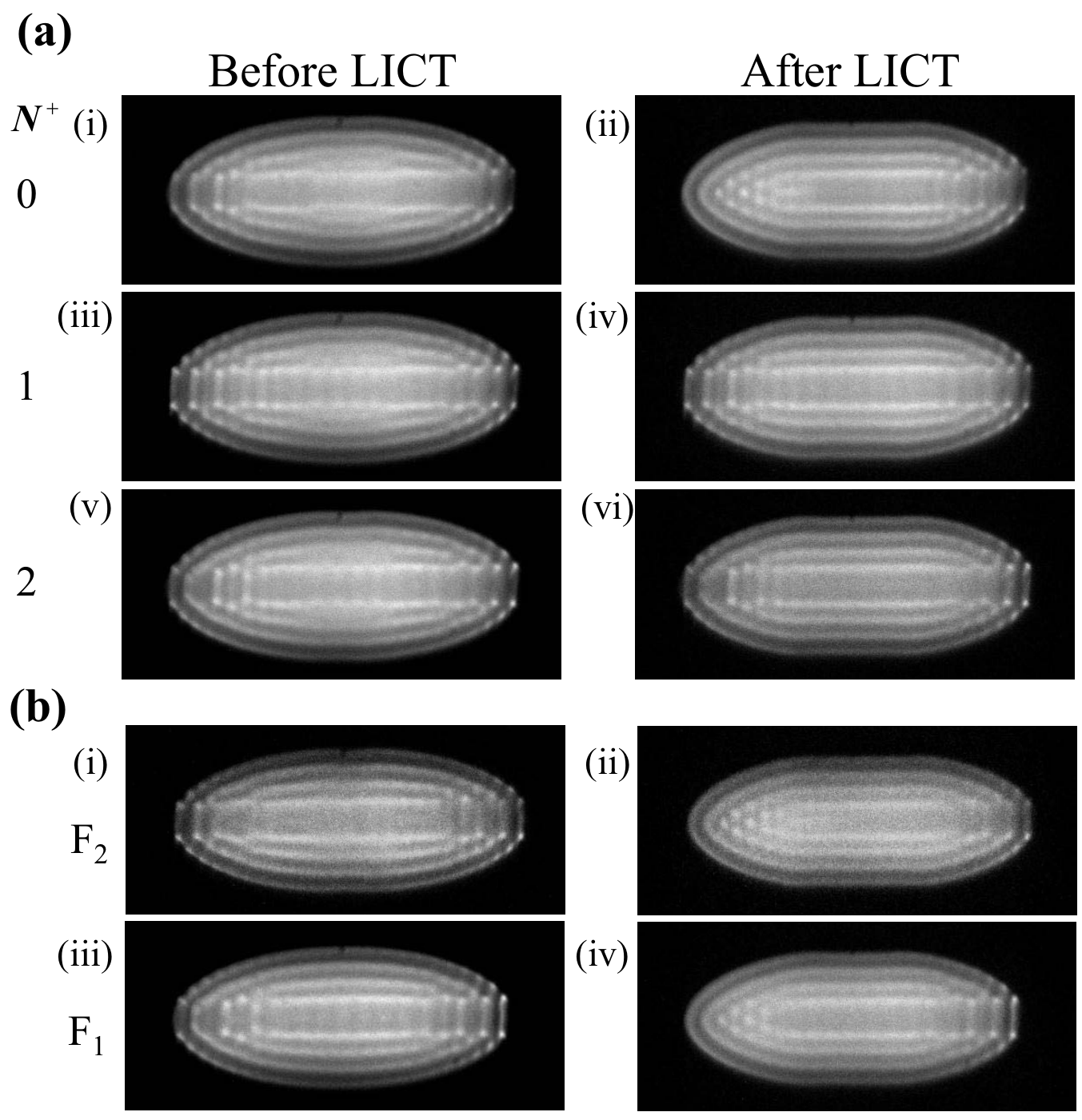,width=0.95\columnwidth}
\caption{\label{ct} LICT experiments to probe the rotational-state populations of N$_2^+$ ions initially prepared in (a) $N^+=0$ and (b) $N^+=3$ after sympathetic cooling. (a) LICT was only observed out of $N^+=0$ demonstrating that the rotational state is preserved during sympathetic cooling. (b) LICT experiments on the $F_1$ and $F_2$ spin-rotation components of N$_2^+$ ions prepared in $N^+=3$, see text for details. }
\end{figure}

Fig. \ref{ct} (a) shows a LICT experiment on N$_2^+$ ions produced in the ground rotational state $N^+=0$. The ions were generated by threshold ionization from the $a''~^1\Sigma^+_g,~v'=0, ~J'=2$ intermediate level excited via the S(0) rovibronic transition from the ground state of neutral N$_2$ (see Fig. \ref{exc} (a)) \cite{hanisco91a}. Using the ionization energies reported in Ref. \cite{mackenzie95a}, the wavenumber of the ionization laser was set 4.0 cm$^{-1}$ above the $N^+=0$ ionization threshold. With this ionization scheme, {\it only} $N^+=0$ ions can be produced for energetic reasons on a weak $\Delta N=-2$ ionizing transition \cite{mackenzie95a}. Fig. \ref{ct} (a) (i) shows a fluorescence image obtained after sympathetically cooling 24$\pm$1 state-selected N$_2^+$ ions in $N^+=0$ into a Coulomb crystal of 925$\pm$25 Ca$^+$ ions. 

The population in $N^+=0$ was probed by LICT immediately after sympathetic cooling by exciting the $X^+~^2\Sigma_{g}^+, \ v^+=0\rightarrow A^+~^2\Pi_{u}, v=2, R_{11}$ transition (see Fig. \ref{exc} (b)) and admitting Ar gas into the chamber. After LICT, the image in Fig. \ref{ct} (a) (ii) was obtained in which $50\pm8$ \% of the N$_2^+$ ions (12$\pm1$ out of 24$\pm1$) have been removed. Averaged over five experiments, we obtain a LICT efficiency of $51\pm6$~\% in agreement with the maximum value of 55\% which can be achieved by the current scheme. We thus conclude that the generation of N$_2^+$ ions is fully state selective (ground-state population $93\pm11$~\%) and that the population is preserved in $N^+=0$ during the sympathetic cooling process. This observation was also confirmed by control experiments showing no evidence of population in the $N^+=1$ and 2 states within our uncertainty limits, see Figs. \ref{ct} (a) (iii)-(vi). These results are in agreement with previous studies \cite{bertelsen06a, koelemeij07a} in which no coupling between the internal and external molecular degrees of freedom was observed during sympathetic cooling. 

Fig. \ref{ct} (b) shows LICT experiments on N$_2^+$ ions prepared in the $N^+=3$ state. Photoionization was carried via the $a''~^1\Sigma^+_g,~v'=0, ~J'=3$ intermediate level by setting the wavenumber of the ionization laser 8.8 cm$^{-1}$ above the $N^+=3$, see Fig. \ref{exc} (a). Figs. \ref{ct} (b) (i)-(iv) show examples for LICT experiments probing the population in the two spin-rotation components $F_1$ and $F_2$ of the $N^+=3$ state. The efficiencies for LICT out of both sub-levels agree within the measurement uncertainties ($27\pm7$ \% and $28\pm7$ \% for $F_1$ and $F_2$, respectively, averaged over five experiments) indicating that both components are produced with equal probability using the current photoionization scheme. The total measured LICT efficiency out of $N^+=3$ amounts to $55\pm9$~\% in agreement with the maximum value of 55\%. Control experiments probing the population in the $N^+=1,2,4$ and 5 states yielded no evidence for population in these levels within the uncertainty limits. In particular, no evidence for population in $N^+=1$ was found indicating that the intensity of the weakly allowed $J'=3\rightarrow N^+=1$ photoionization channel is considerably reduced compared to $J'=3\rightarrow N^+=3$ in line with the results of previous rotationally resolved photoelectron spectroscopic studies \cite{mackenzie95a}. 

\begin{figure}[t]
\epsfig{file=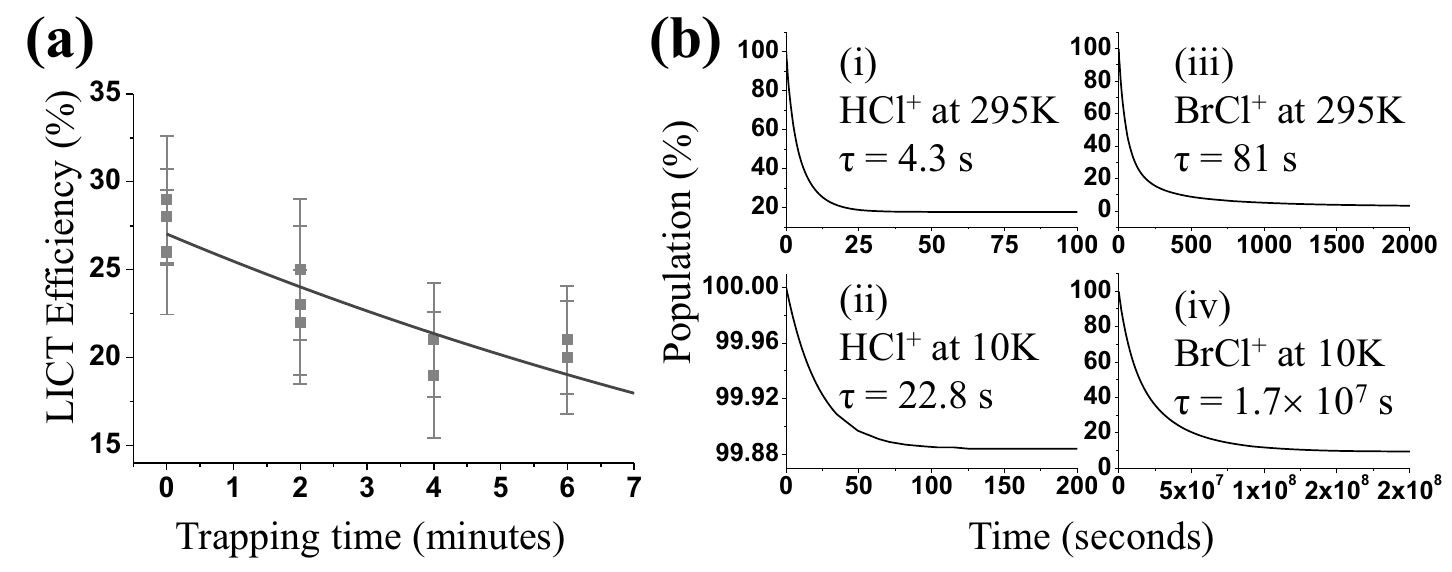,width=\columnwidth}
\caption{\label{lt} (a) Decrease of the population of N$_2^+$ ions in $N^+=3, ~F_2$ as a function of the trapping time. Squares: experimental data, solid line: fit to a pseudo-first order rate law yielding a lifetime $\tau=947\pm176$~s, see text. (b) Theoretical population-decay curves and radiative lifetimes $\tau$ for selected polar diatomic ions initially prepared in the rotational ground state under the influence of black-body radiation.}
\end{figure}

The lifetime of the initially prepared $N^+=3, ~F_2$ state was probed by introducing a variable time delay between loading the crystal with N$_2^+$ ions and LICT, see Fig. \ref{lt} (a). Over a timescale of 6 minutes, we observe a decrease of the number of state-selected N$_2^+$ ions $N_{\text{N}_2^+}$ commensurate with a pseudo-first order rate law $N_{\text{N}_2^+}(t)/N_{\text{N}_2^+}(t=0)=\exp\big\{-\tau^{-1}t\big\}$. The lifetime $\tau$ was determined to be $\tau=947\pm 176$~s. We attribute the loss of population out of $N^+=3$ to inelastic and reactive collisions with background gas molecules \cite{tong10b}. 

The method presented here can be used for the generation of a wide range of state-selected ions provided that a state-selective photoionization scheme can be devised on the basis of, e.g., high-resolution photoelectron spectroscopy or theoretical considerations \cite{willitsch05b}. This requirement is met for a variety of di- and polyatomic molecules such as H$_2$, O$_2$, CO, HCl, NH$_3$, C$_2$H$_4$, H$_2$O and C$_6$H$_6$ among others \cite{willitsch05b}. This list also includes precursors of polar ions for which the rotational-state lifetimes in the trap are expected to be limited by dipole coupling to the ambient black-body radiation field \cite{bertelsen06a, koelemeij07a}. For polar molecules, the timescales for the redistribution of the rotational population are dictated by the magnitude of the molecular dipole moment, the rotational constants and the BBR intensity \cite{hoekstra07a}. We have modeled the population redistribution caused by BBR for selected, prototypical polar ions using a rate-equation model for radiative coupling \cite{tong10b}. For a molecular ion with a small dipole moment $\mu$ and a small rotational constant $B_e$ such BrCl$^+$ ($\mu=0.26$~D, $B_e=0.17$~cm$^{-1}$ \footnote{Calculated at the UMP2/cc-pVTZ level of theory.}), we predict that the relevant optical transitions lie in spectral regions with low BBR intensity at room temperature. The initially prepared state can be retained on timescales exceeding minutes which can be lengthened by several orders of magnitude by lowering the BBR intensity in a cryogenic environment (see Fig. \ref{lt} (b) (iii), (iv)). For a strongly polar ion with a large rotational constant such as HCl$^+$ ($\mu=1.64$~D, $B_e=9.96$~cm$^{-1}$ \cite{huber79a}), we predict a lifetime of the rotational ground state of only 4.3 s (Fig. \ref{lt} (b) (i)). However, at a BBR temperature of 10 K the population is almost completely confined to the lowest rotational level in thermal equilibrium so that ions prepared in the ground state can be retained for extended periods of time (Fig. \ref{lt} (b) (ii)). Excited rotational levels, on the other hand, decay on timescales of tens of seconds \cite{tong10b}. 

Our method can be further developed in several directions. Selected rotational levels in vibrationally excited states can be produced by taking advantage of diagonal Franck-Condon factors in the photoionization of suitable Rydberg states \cite{conaway87a}. Additionally, molecular ions in selected spin-rotation and even hyperfine states could be generated using extremely narrow-bandwidth-laser sources for threshold ionization \cite{paul09a}. We expect that the method presented here will be particularly useful to prepare state-selected, translationally cold ions for cold ion-molecule collision experiments \cite{willitsch08a} and for single-ion frequency-metrology studies \cite{schiller05a} as well as to initialize molecular-ion qubits.

\bibliographystyle{apsrev4-1}
\bibliography{/Users/stefanwillitsch/Documents/docs/bib_file/main}

\end{document}